\documentclass[10pt, paper=a4, UKenglish]{article}
\usepackage{graphicx}
\usepackage{xspace}
\def\Title#1{\begin{center} {\Large #1 } \end{center}}
\def\Author#1{\begin{center}{ \sc #1} \end{center}}
\def\Address#1{\begin{center}{ \it #1} \end{center}}

\newcommand\pubblock{\rightline{\begin{tabular}{l} Proceedings of the CTD 2023\\ \pubnumber\\
         \pubdate  \end{tabular}}}

\newenvironment{Abstract}{\begin{quotation} \begin{center} 
             \large ABSTRACT \end{center}\bigskip 
      \begin{center}\begin{large}}{\end{large}\end{center} \end{quotation}}

\newenvironment{Presented}{\begin{quotation} \begin{center} 
             PRESENTED AT\end{center}\bigskip 
      \begin{center}\begin{large}}{\end{large}\end{center} \end{quotation}}

\def\Acknowledgements{\bigskip  \bigskip \begin{center} \begin{large}
      \bf ACKNOWLEDGEMENTS \end{large}\end{center}}

\def\beq{\begin{equation}}
\def\eeq#1{\label{#1}\end{equation}}
\def\eeqn{\end{equation}}

\def\beqa{\begin{eqnarray}}
\def\eeqa#1{\label{#1}\end{eqnarray}}
\def\eeqan{\end{eqnarray}}

\let\bar=\overbar

\def\Dslash{\not{\hbox{\kern-4pt $D$}}}
\def\dslash{\not{\hbox{\kern-2pt $\del$}}}

\def\msb{{\bar{\ssstyle M \kern -1pt S}}}

\def\pt{\ensuremath{p_\mathrm{T}}\xspace}

\usepackage{color}
\usepackage{lineno}
\usepackage{subfig}
\usepackage{hyperref}

\newcommand\pubnumber{PROC-CTD2023-24}

\newcommand\pubdate{\today}

\def\ucsd{
Department of Physics \\
UC San Diego, USA}

\def\ufl{
Department of Physics \\
University of Florida, USA}

\def\cornell{
Department of Physics \\
Cornell University, USA}

\def\princeton{
Department of Physics \\
Princeton University, USA}

\newcommand{\conference}{Connecting the Dots Workshop (CTD 2023)\\
October 10-13, 2023}

\usepackage{fancyhdr}
\definecolor{mygrey}{RGB}{105,105,105}
\begin{document}


\large
\begin{titlepage}
\pubblock

\vfill
\Title{Improving tracking algorithms with machine learning: a case for line-segment tracking at the High Luminosity LHC}
\vfill

\Author{Jonathan Guiang, Slava Krutelyov, Manos Vourliotis, Yanxi Gu, Avi Yagil, Balaji Venkat Sathia Narayanan, Matevz Tadel}
\Address{\ucsd}
\Author{Philip Chang, Mayra Silva}
\Address{\ufl}
\Author{Gavin Niendorf, Peter Wittich, Tres Reid}
\Address{\cornell}
\Author{Peter Elmer}
\Address{\princeton}
\vfill
\Address{On behalf of the CMS Collaboration}
\vfill
\clearpage

\begin{Abstract}
In this work, we present a study on ways that tracking algorithms can be improved with machine learning (ML). We base this study on a line-segment-based tracking (LST) algorithm that we have designed to be naturally parallelized and vectorized in order to efficiently run on modern processors. LST has been developed specifically for the Compact Muon Solenoid (CMS) Experiment at the LHC, towards the High Luminosity LHC (HL-LHC) upgrade. Moreover, we have already shown excellent efficiency and performance results as we iteratively improve LST, leveraging a full simulation of the CMS detector. At the same time, promising deep-learning-based tracking algorithms, such as Graph Neural Networks (GNNs), are being pioneered on the simplified TrackML dataset. These results suggest that parts of LST could be improved or replaced by ML. Thus, a thorough, step-by-step investigation of exactly how and where ML can be utilized, while still meeting realistic HL-LHC performance and efficiency constraints, is implemented as follows. First, a lightweight neural network is used to replace and improve upon explicitly defined track quality selections. This neural network is shown to be highly efficient and robust to displaced tracks while having little-to-no impact on the runtime of LST. These results clearly establish that ML can be used to improve LST without penalty. Next, exploratory studies of GNN track-building algorithms are described. In particular, low-level track objects from LST are considered as nodes in a graph, where edges represent higher-level objects or even entire track candidates. Then, an edge-classifier GNN is trained, and the efficiency of the resultant edge scores is compared with that of the existing LST track quality selections. These GNN studies provide insights into the practicality and performance of using more ambitious and complex ML algorithms for HL-LHC tracking at the CMS Experiment.
\end{Abstract}

\vfill

\begin{Presented}
\conference
\end{Presented}
\vfill
\end{titlepage}
\def\thefootnote{\fnsymbol{footnote}}
\setcounter{footnote}{0}
%

\normalsize 


\section{Introduction}
\label{sec:intro}
\begin{figure}[!htb]
  \centering
  \subfloat{\includegraphics[width=0.45\linewidth]{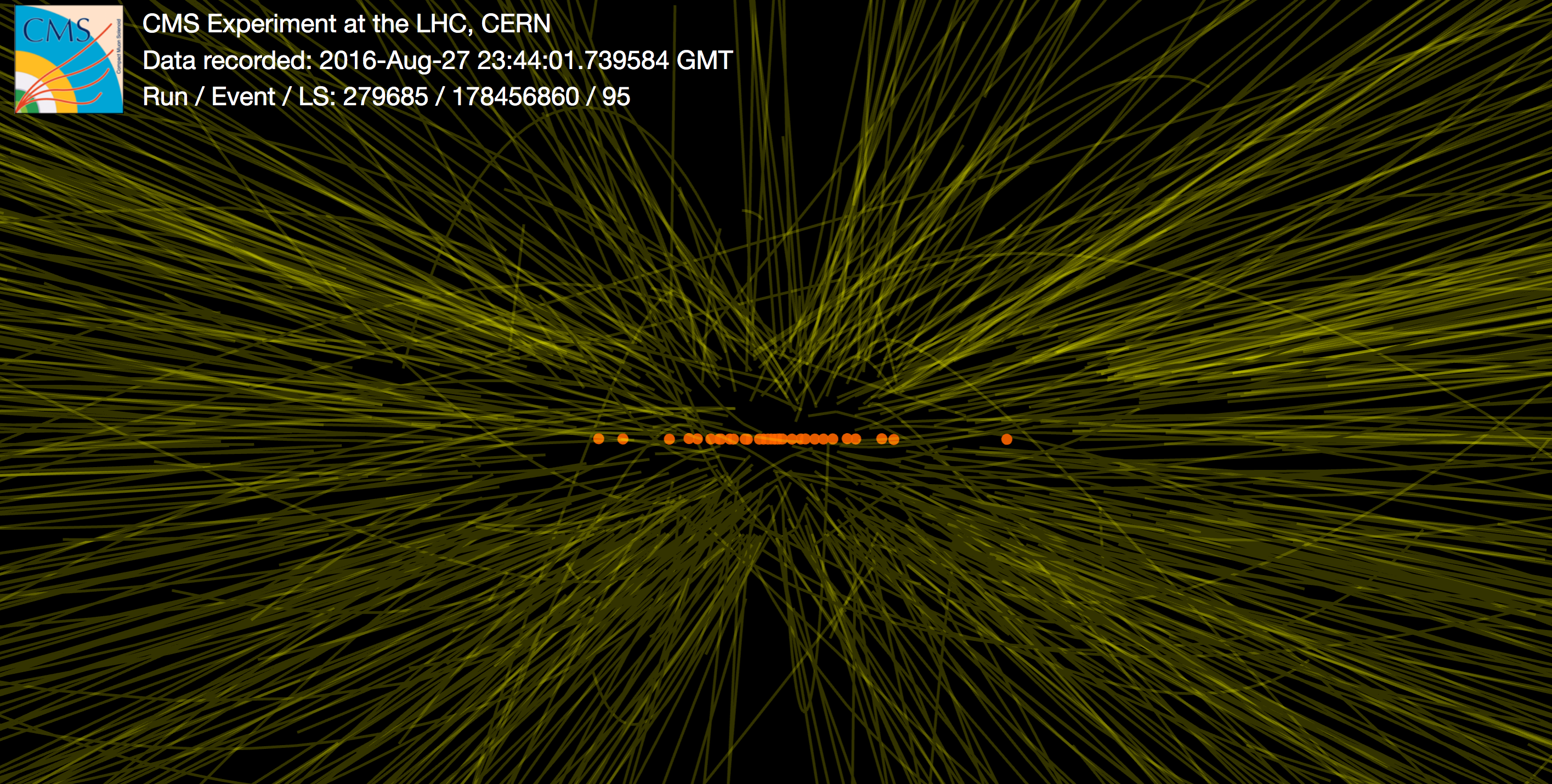}}
  \qquad
  \subfloat{\includegraphics[width=0.45\linewidth]{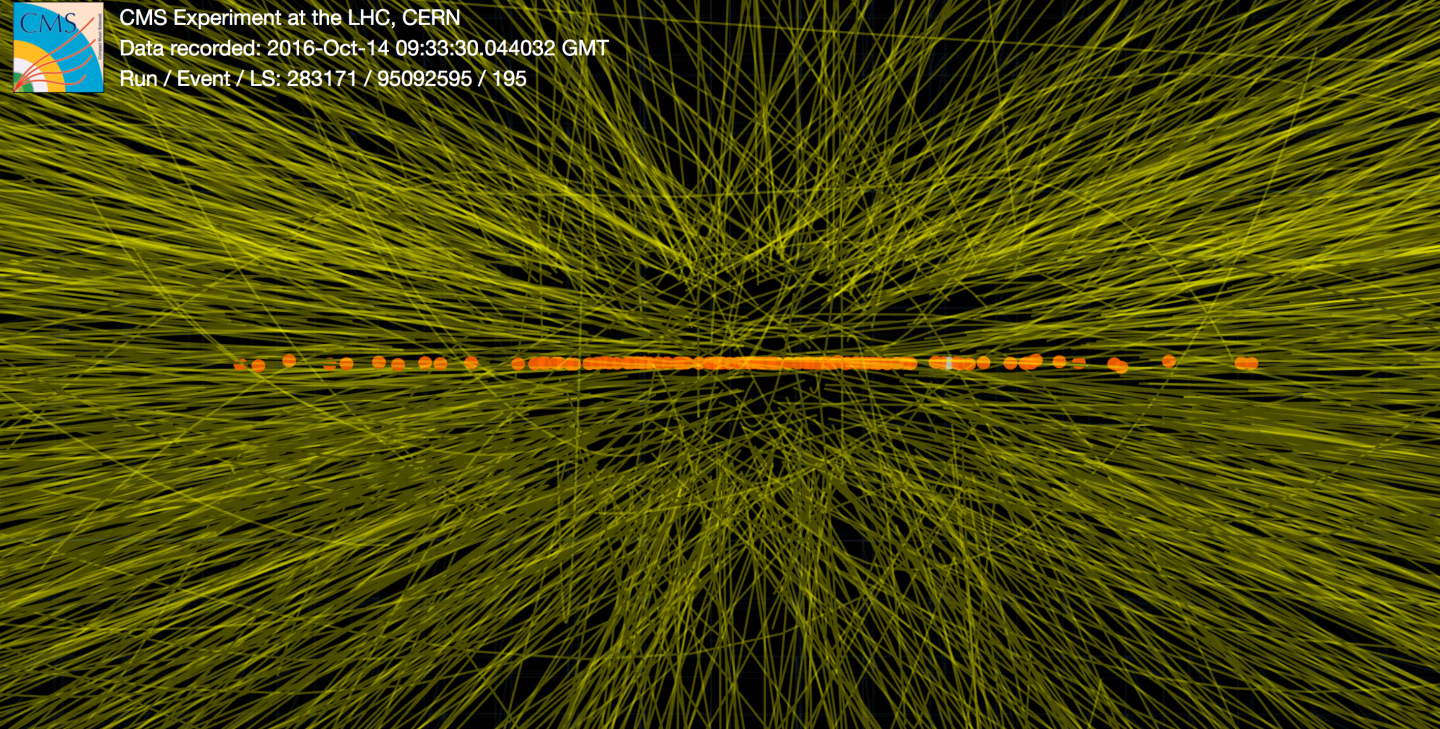}}
  \caption{
      A collision event with standard pile-up (left) recorded at CMS in 2016 is shown next to an event with HL-LHC-like pile-up (right) recorded at CMS in the same year~\cite{cite:2016NormPU, cite:2016HighPU}. 
      The dots are proton-proton collisions and the thin lines are the reconstructed particle tracks.
  }
  \label{fig:hl-lhc-pileup}
\end{figure}

At the LHC, a significant effort is underway to prepare for the ``High Luminosity'' LHC (HL-LHC) era~\cite{cite:hl-lhc}, wherein the number of concurrent proton-proton collisions (``pile-up'') is planned to be increased by an order of magnitude (Fig.~\ref{fig:hl-lhc-pileup}), enabling the observation of rare processes. 
Among the many challenges that this brings, charged particle tracking will be one of the most difficult: with massively increased pile-up, many thousands of spurious hits are registered in the tracker for each interesting proton-proton collision event.
HL-LHC particle tracking algorithms must therefore efficiently identify real tracks amongst more noise, and they must also scale well with increased tracker occupancy. 
Moreover, while CMS will acquire more computational resources, current projections indicate that they will not be sufficient to meet expected HL-LHC demands~\cite{cite:CMS-NOTE-2022-008}.
Most current algorithms, however, are inherently sequential and thus scale poorly, so innovative solutions are required in order to ensure the success of the HL-LHC program at CMS. 

We have proposed a solution for HL-LHC track-finding at CMS in previous work: the line-segment tracking (LST) algorithm, designed to be naturally parallelizable~\cite{cite:ctd22}. 
In particular, LST builds ``segments'' of tracks using only information local to that segment, such that they may all be built in parallel. 
The algorithm starts in the outer tracker, which will have modules, referred to as ``\pt-modules,'' that have two silicon sensors separated by a few millimeters~\cite{cite:CERN-LHCC-2017-009}. 
Two variants of \pt-module will be implemented: one with two strip sensors (2S) and another with one macro-pixel sensor and one strip sensor (PS). 
Two hits in each \pt-module are considered as track segments called ``Mini-Doublets'' (MDs), and they are required to have $\pt > 0.8$ GeV.
Next, two MDs in nearby modules are considered as longer track segments called ``line segments'' (LSs). 
LSs may only be built across adjacent tracker modules that fulfill some geometric criteria and are required to have MDs with consistent \pt estimates. 
Step by step, progressively longer track objects are constructed: LSs sharing a common MD are considered as ``triplets'' (T3s), then T3s sharing a common MD are considered as ``quintuplets'' (T5s). 
For each track segment, pre-defined quality selections are applied, enforcing consistency with a continuous, helical trajectory. 
Finally, pixel seeds (pLSs or initial iteration seeds) are obtained from an upstream iteration of the CMS track-finding workflow that runs on the hits from the inner tracker. 
These pLSs are matched to the outer tracker LST track segments to form four mutually exclusive types of track candidates (TCs), with preference given to the longest tracks. 
Any ``unused'' pLSs --- pLSs that are not matched to a T5 or T3 --- are still written to the final list of TCs in order to maintain efficiency beyond the extend of the outer tracker. 
The list of TCs as well as all of the LST steps are summarized in Table~\ref{tab:lst}.

\begin{figure}[!htb]
  \centering
  \subfloat{\includegraphics[width=0.45\linewidth]{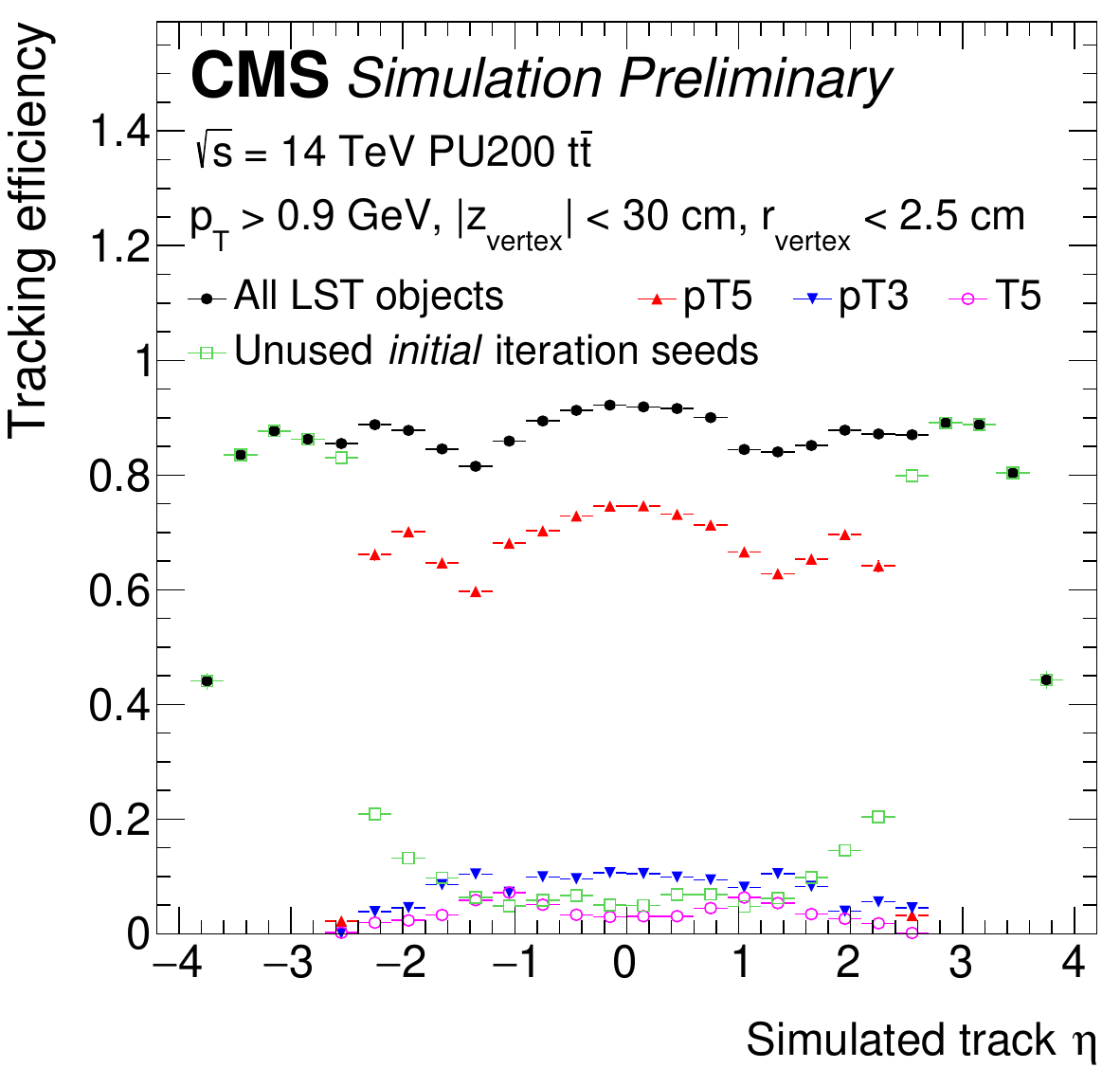}}
  \qquad
  \subfloat{\includegraphics[width=0.45\linewidth]{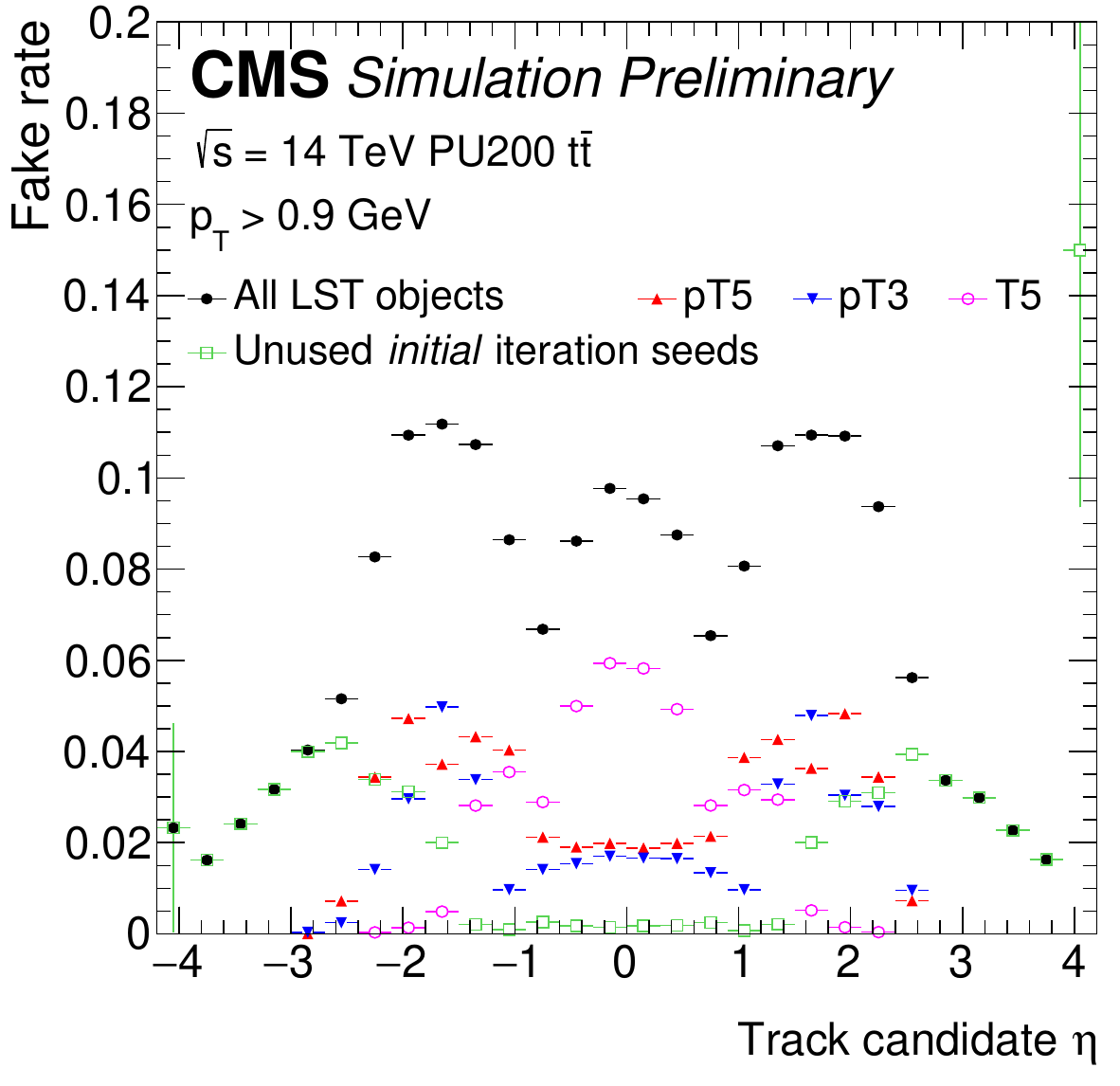}}
  \caption{
      The LST track-finding efficiency (left) and fake rate (right) are plotted as a function of the pseudorapidity $\eta$ of the simulated track and track candidate respectively. 
      For both plots, five histograms are overlayed: all track candidates (black), pT5s (red), pT3s (blue), T5s (magenta), and pLSs that are not used in a pT5 or pT3 (green). 
      On the left, it can be seen that T5s comprise the majority of the LST track-finding efficiency, either as pT5s or T5s. 
      On the right, it can also be seen that the T5s make up the bulk of the LST fake rate in the barrel region. 
  }
  \label{fig:lst-performance}
\end{figure}

The latest performance metrics for LST are shown in Fig.~\ref{fig:lst-performance} for simulated $t\bar{t}$ events with HL-LHC conditions. 
The simulated tracks, or collections of true hits left behind by each simulated particle, are used to evaluate the overall tracking performance.
In particular, it is clear that T5s contribute the largest fraction of the overall efficiency, as they enter the TCs as pT5s or T5s. 
The efficiency is defined as the fraction of simulated tracks that are matched to a TC, where a TC is considered as matching a simulated track if more than 75\% of the hits in the TC belong to that simulated track. 
At the same time, T5s contribute the largest fraction of the overall fake rate, or the rate at which T5s that are not matched to a simulated track (called ``fake'' T5s) are selected by LST as a TC. 
The T5 fake rate is particularly high in the ``barrel'' region of CMS, defined by $|\eta| < 2.5$ where $\eta$ is the pseudorapidity. 
There is therefore an opportunity in the T5-building step to improve the final performance of the LST algorithm: by building more real T5s or fewer fake T5s, the downstream TC efficiency or fake rate would be similarly improved. 
In order to achieve this, we focus on replacing pre-defined selections with a lightweight deep neural network (DNN) that might use the same variables better.

\begin{table}[!htb]
    \begin{center}
        \begin{tabular}{c|c|p{6cm}}
            \hline
            \hline
            Step & Track segment       & Description                    \\
            \hline
            0 & Pixel seeds (pLSs)  & Track segments from the inner tracker \\
            1 & Mini-doublets (MDs) & Two hits in a \pt-module       \\
            2 & Line segments (LSs) & Two MDs in nearby modules      \\
            3 & Triplets (T3s)      & Two LSs that share a common MD \\
            4 & Quintuplets (T5s)   & Two T3s that share a common MD \\
            \hline
            \hline
            Step & Track candidate     & Description                                             \\
            \hline
            5 & pT5s & T5s matched to a pixel seed                                            \\
            6 & T5s  & T5s that are not matched to a pixel seed                               \\
            7 & pT3s & T3s that are not in a pT5, but are matched to a pixel seed             \\
            8 & pLS  & Pixel seeds that are not already in a pT5 or pT3 \\
            \hline
            \hline
        \end{tabular}
        \caption{
            The steps of the LST algorithm are shown in order of execution, starting with track-segment building (Steps 1 to 4) followed by track-candidate selection (Steps 5 to 8). 
            Step 0 is performed by a preceding iteration of the CMS track-finding algorithm. 
            Each LST step is implemented as a separate kernel, where the track objects of interest are built in parallel.
        }
        \label{tab:lst}
    \end{center}
\end{table}

\section{T5-DNN training}
\label{sec:training}
\begin{figure}[!htb]
  \centering
  \subfloat[]{\includegraphics[width=0.35\linewidth]{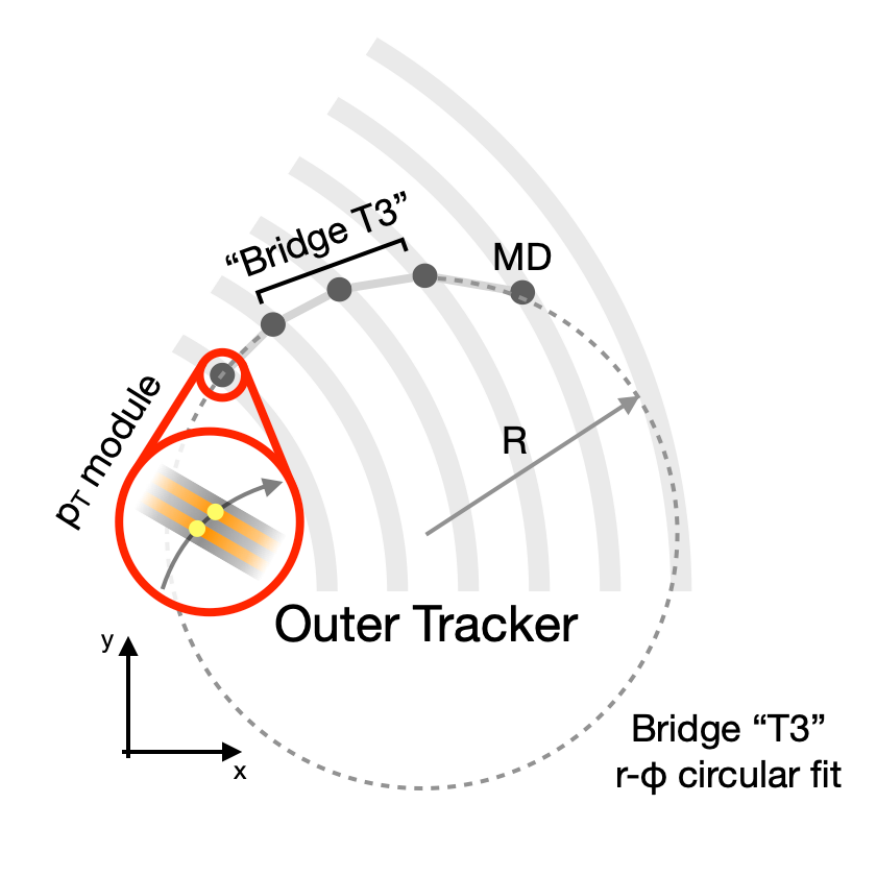}\label{fig:t5-anatomy}}
  \qquad
  \subfloat[]{\includegraphics[width=0.55\linewidth]{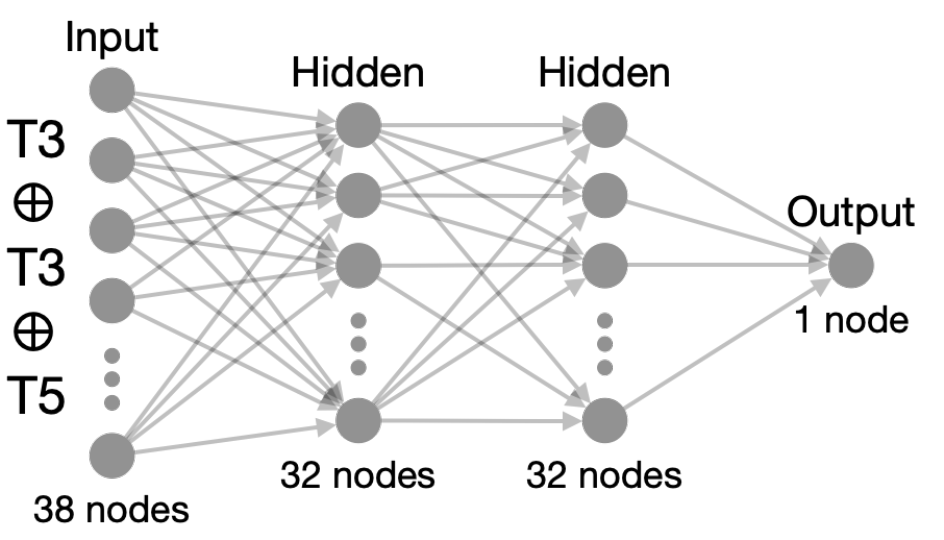}\label{fig:dnn-arch}}
  \caption{
      The components of a T5 are depicted, showing the ``bridge T3,'' the r-$\phi$ circular fit, and a MD (a) next to a diagram of the T5-DNN architecture (b).
  }\end{figure}

The DNN in this work was trained on a larger set of T5s than what LST produced by default. 
Specifically, a set of predefined selections on a custom heuristic correlated to the quality of a circular fit to the T5 candidate hits in the $r$-$\phi$ plane were removed under the hypothesis that the DNN could better capture this information. 
LST was run with this modification on 175 $t\bar{t}$ events with HL-LHC pile-up. 
This produced a dataset of 2.1 million T5s for training, of which approximately 40\% are real T5s. 
The dataset was shuffled and split: 80\% of the data was reserved for training, with the remaining 20\% used for testing

Since a T5 is composed of two T3s that shared a common MD, both T5-level and T3-level features are used as inputs to the DNN. 
In addition, a ``Bridge T3'' (Fig.~\ref{fig:t5-anatomy}) is defined as the three connected MDs at the center of the T5. 
This provides an additional subset of hits which the DNN can, in principle, compare to the other T3s for consistency. 
The coordinates for an ``anchor'' hit, one of the two hits in the MD, in the T5 are also provided. 
For PS \pt-modules, the hit in the macro-pixel sensor is taken as the anchor hit, while for 2S \pt-modules, the hit in the innermost (closest to the beamline) strip sensor is taken as the anchor hit. 
In total, 38 inputs are provided: the \pt estimate and radius of a simple circle fit for each of the two T3s; the $(r,\phi,z)$ coordinates, pseudorapidity $\eta$, and layer of each anchor hit; the \pt, $\eta$, and $\phi$ of the T5; and the radius of a simple circle fit for the bridge T3. 
Since the T5-DNN ultimately classifies real versus fake T5s, it can be formally defined as a binary classification task. 
As such, Binary Cross Entropy, which is most suitable for binary classification tasks, was selected as the loss function.

\begin{figure}[!htb]
  \centering
  \includegraphics[width=0.65\linewidth]{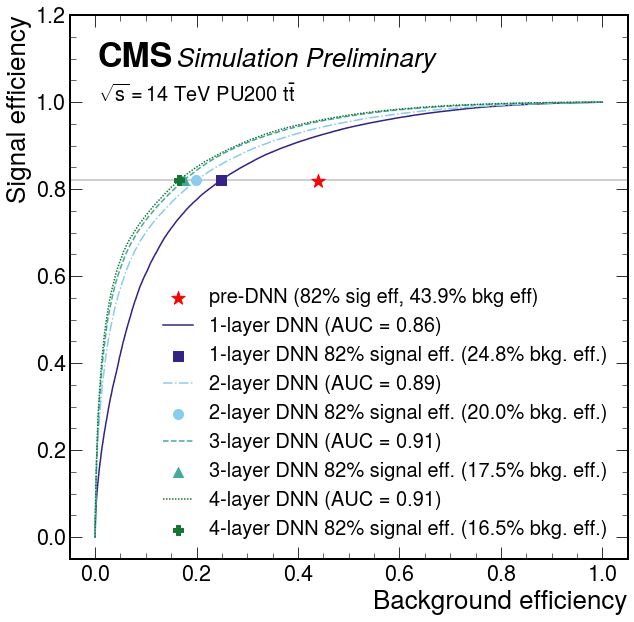}
  \caption{
    The ROC curves for a 1-layer, 2-layer, 3-layer, and 4-layer DNN are plotted, with the signal efficiency on the y-axis and the background efficiency, or fake rate, on the x-axis. 
    Starting with 1 layer, each additional hidden layer gives an additional 3\% improvement in the fake rate with respect to pre-DNN LST plotted as a red star. 
    The 2-layer DNN was ultimately selected for this work.
  }
  \label{fig:roc}
\end{figure}

A simple architecture (Fig.~\ref{fig:dnn-arch}) for the T5-DNN was selected: a two-layer deep neural network, where each hidden layer has only 32 nodes. 
Since the DNN inference needs to be run for every candidate T5 considered by the LST algorithm, a smaller architecture was preferred. 
Thus, the two-layer DNN was selected as a balance of performance and computational complexity.
However, deeper neural networks yielded only marginal improvement over the two-layer DNN (Fig.~\ref{fig:roc}): a 3-layer DNN has a fake rate of 17.5\% and a 4-layer DNN has a fake rate of 16.5\%, while the 2-layer DNN has a fake rate of 20\% at the same signal efficiency. 
Importantly, the T5 DNN is applied only at the T5-building step, so while a fake rate around 20\% is relatively large, it will be further reduced downstream in LST. 
Matching with the pixel seeds, additional quality cuts, and other steps like duplicate removal all have a non-trivial effect on the final performance. 

\begin{figure}[!htb]
  \centering
  \subfloat[]{\includegraphics[width=0.45\linewidth]{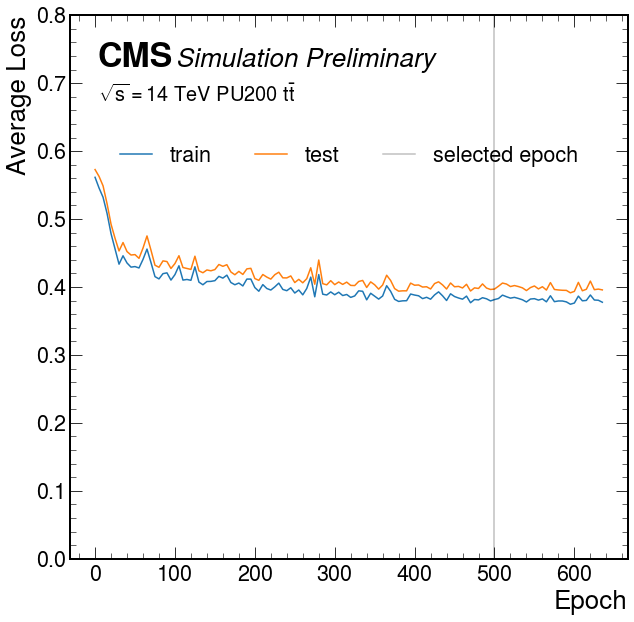}\label{fig:history}}
  \qquad
  \subfloat[]{\includegraphics[width=0.45\linewidth]{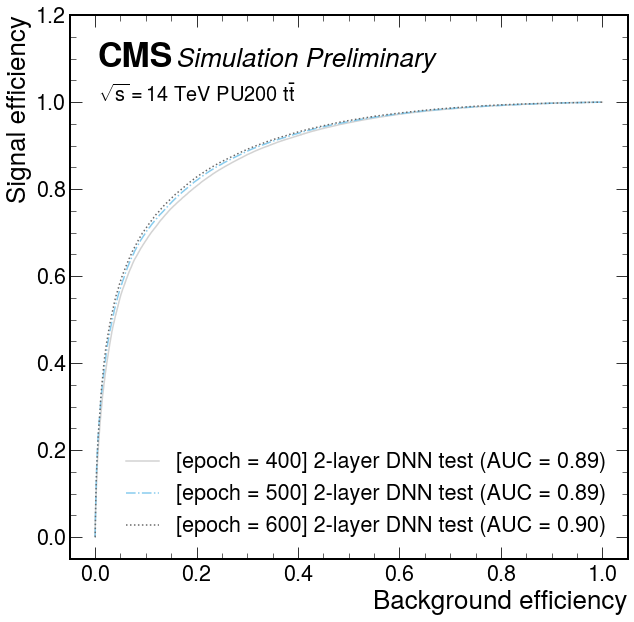}\label{fig:roc-epochs}}
  \caption{
      The average loss after each epoch is plotted (a) next to the Receiver Operating Characteristic (ROC) curves for the model after epoch 400, 500, and 600 (b). 
      For the ROC curves, the signal efficiency is plotted on the y-axis, while the background efficiency, or fake rate, is plotted on the x-axis. 
      It can be seen from these plots that the loss plateaus after epoch 400, and that the choice of model after that epoch is arbitrary, as the model performance is identical in epochs 400, 500, and 600. 
  }
\end{figure}

The T5-DNN was trained for around 630 epochs (Fig.~\ref{fig:history}), but the minimization of the loss plateaus after 400 epochs. 
The training after the 500th epoch was selected for this work, though the selection was arbitrary as the performance of the models at epoch 400, 500, and 600 were effectively identical (Fig.~\ref{fig:roc-epochs}). 
Overall, the T5-DNN reduces the fake rate for classifying T5s by a factor of two (Fig.~\ref{fig:roc}). 
The full impact of the T5-DNN on LST, after all of the downstream steps mentioned previously, is assessed in the next section.

\section{T5-DNN performance}
\label{sec:performance}
\begin{figure}[!htb]
  \centering
  \includegraphics[width=0.85\linewidth]{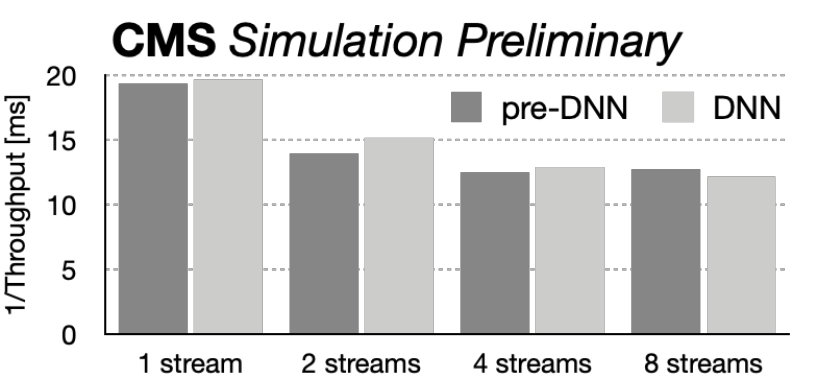}
  \caption{The 2022 CPU usage projections are shown, where it can be seen that R\&D efforts must be undertaken alongside resource increases in order to meet the demand in 2030 when the HL-LHC is planned to start data taking~\cite{cite:CMS-NOTE-2022-008}.}
  \label{fig:streams-vs-throughput}
\end{figure}

After training, the T5-DNN was integrated into LST through the simplest implementation: the weights are hard-coded as C++ arrays, and the DNN matrix multiplication is done through nested for-loops. 
The inference is run for each T5 considered by LST, and a working point was selected to match the original LST efficiency. 
Despite the simplicity of the implementation, no significant impact on the standalone LST runtime is observed (Table~\ref{tab:throughput}). 
This is consistent for 1, 2, 4, and 8 CUDA streams (Fig.~\ref{fig:streams-vs-throughput}), where each stream handles event-level parallelization opportunistically.

\begin{table}[!htb]
    \begin{center}
        \begin{tabular}{l|ccc}
            \hline
            \hline
            Trial   & T5 [ms]         & 1/Throughput [ms] & N streams \\
            \hline
            pre-DNN & $3.37 \pm 0.13$ &  $28.4 \pm 1.5$   & 1         \\
            DNN     & $3.39 \pm 0.07$ &  $28.7 \pm 1.1$   & 1         \\
            \hline
            \hline
        \end{tabular}
        \caption{
            The runtime for the T5-building stage and the inverse of the overall throughput are tabulated in milliseconds. 
            The error shown is one standard deviation for ten trials.
        }
        \label{tab:throughput}
    \end{center}
\end{table}

\begin{figure}[!htb]
  \centering
  \subfloat{\includegraphics[width=0.45\linewidth]{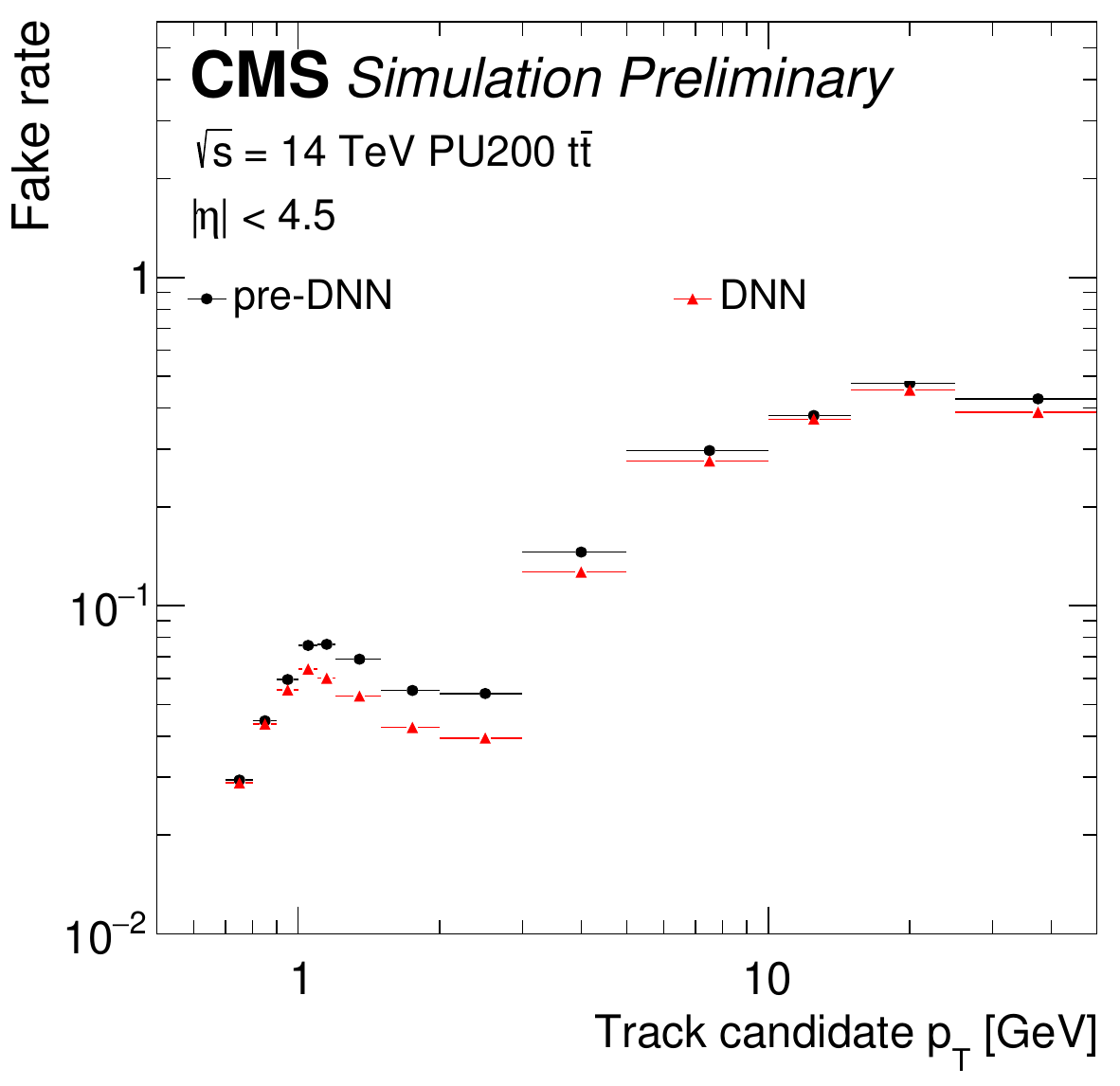}}
  \qquad
  \subfloat{\includegraphics[width=0.45\linewidth]{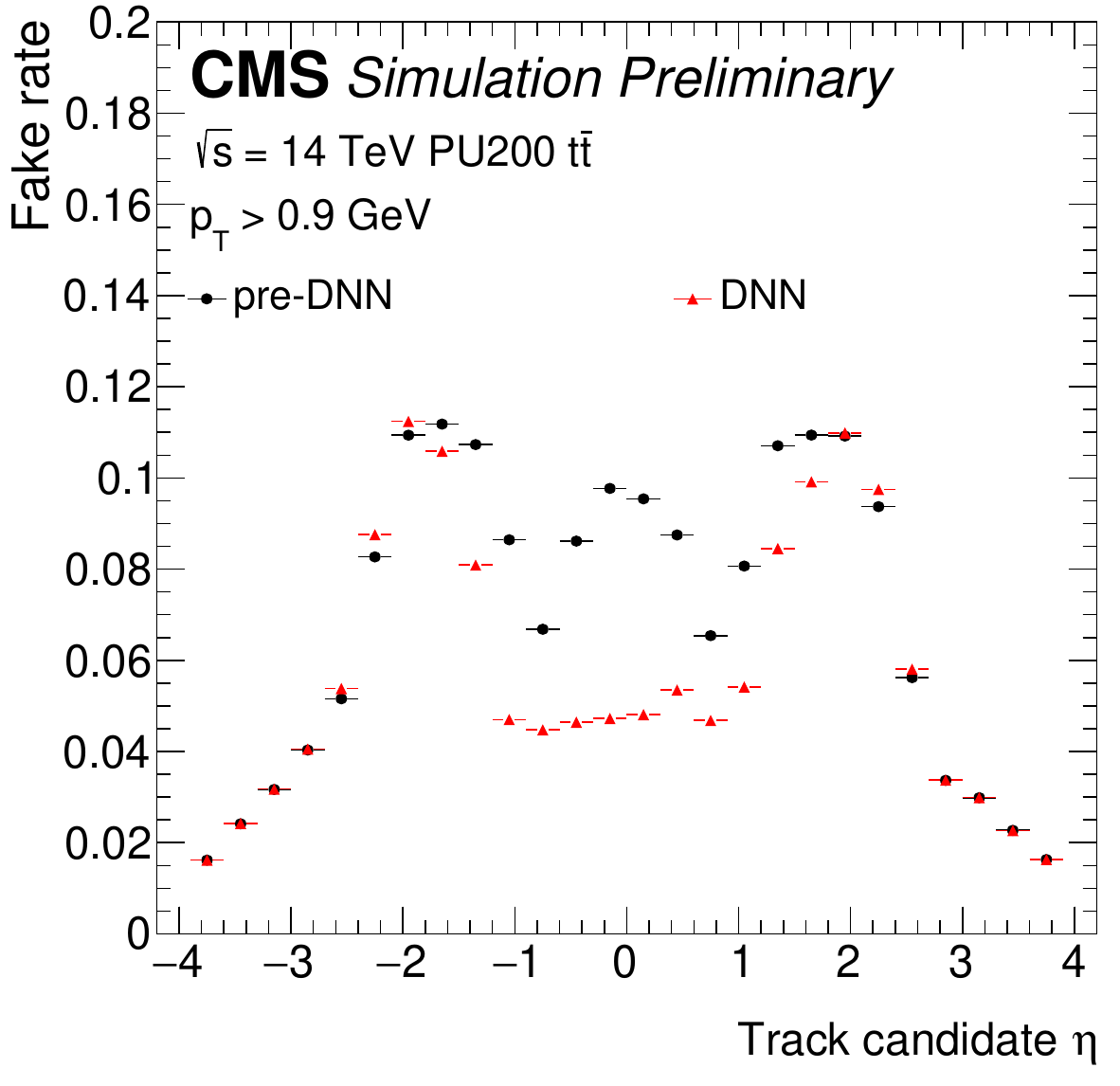}}
  \caption{
      The LST fake rate for all TCs is plotted as a function of \pt (left) and $\eta$ (right). 
      Notably, there is a 40\% reduction in the fake rate in the barrel, where the T5 fake rate was previously dominant. 
  }
  \label{fig:comparison-fr}
\end{figure}

\begin{figure}[!htb]
  \centering
  \subfloat{\includegraphics[width=0.45\linewidth]{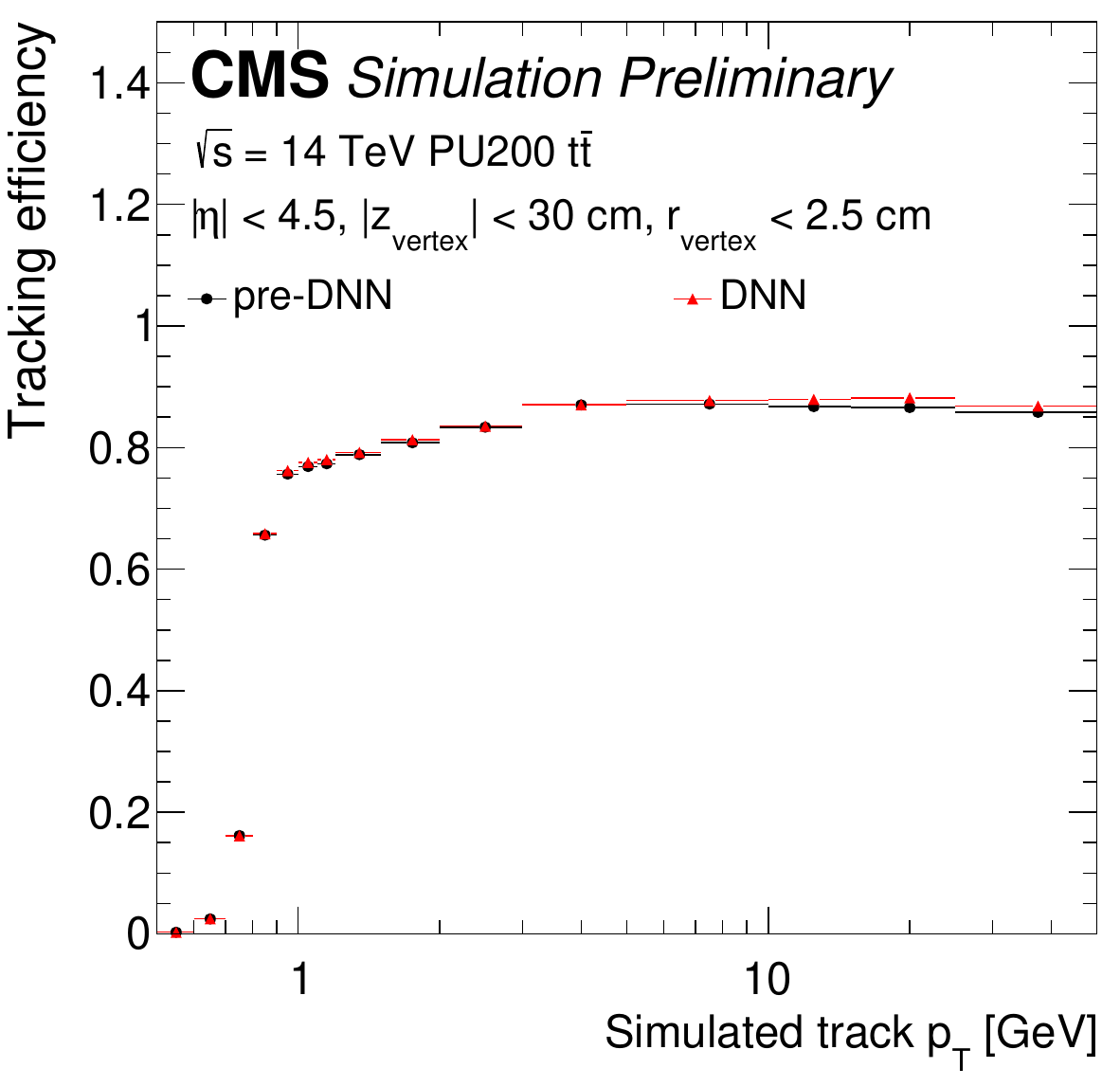}}
  \qquad
  \subfloat{\includegraphics[width=0.45\linewidth]{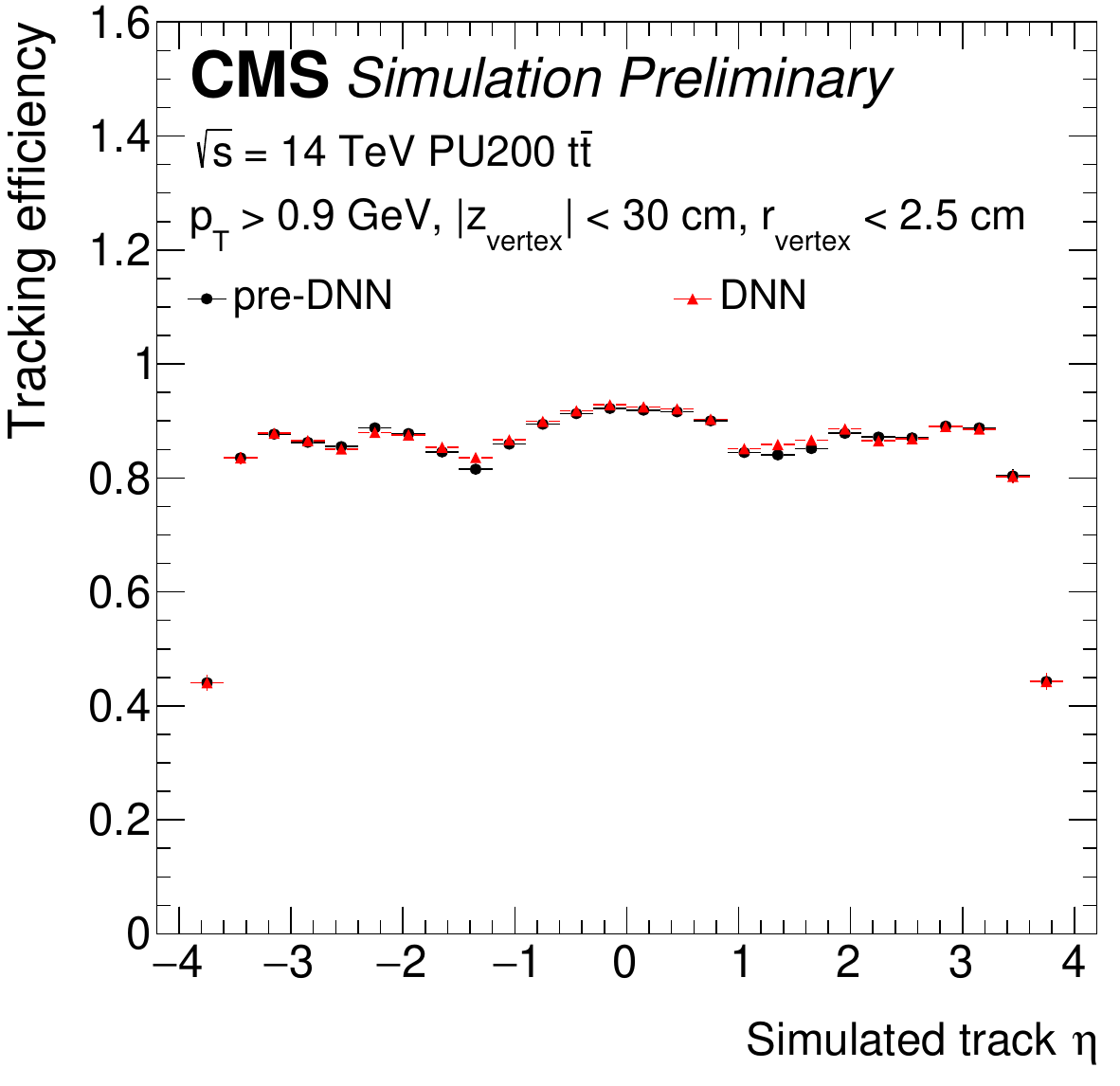}}
  \caption{
      The LST efficiency for all TCs is plotted as a function of \pt (left) and $\eta$ (right). 
      The working point for the DNN was selected to match the efficiency of LST, and it is clear that no efficiency is lost.
  }
  \label{fig:comparison-eff}
\end{figure}

\begin{figure}[!htb]
  \centering
  \subfloat[]{\includegraphics[width=0.45\linewidth]{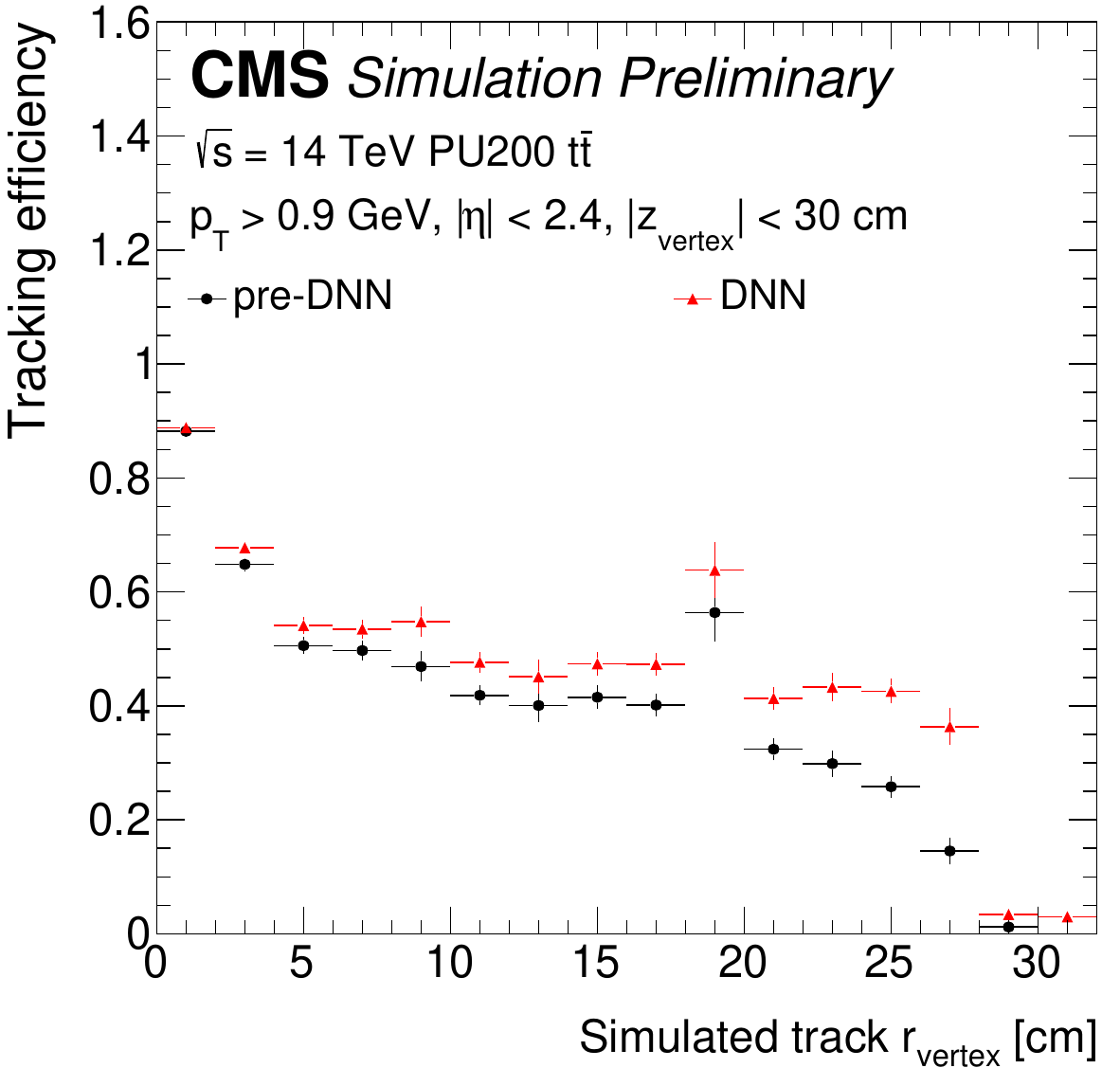}\label{fig:comparison-rvertex-ttbar}}
  \qquad
  \subfloat[]{\includegraphics[width=0.45\linewidth]{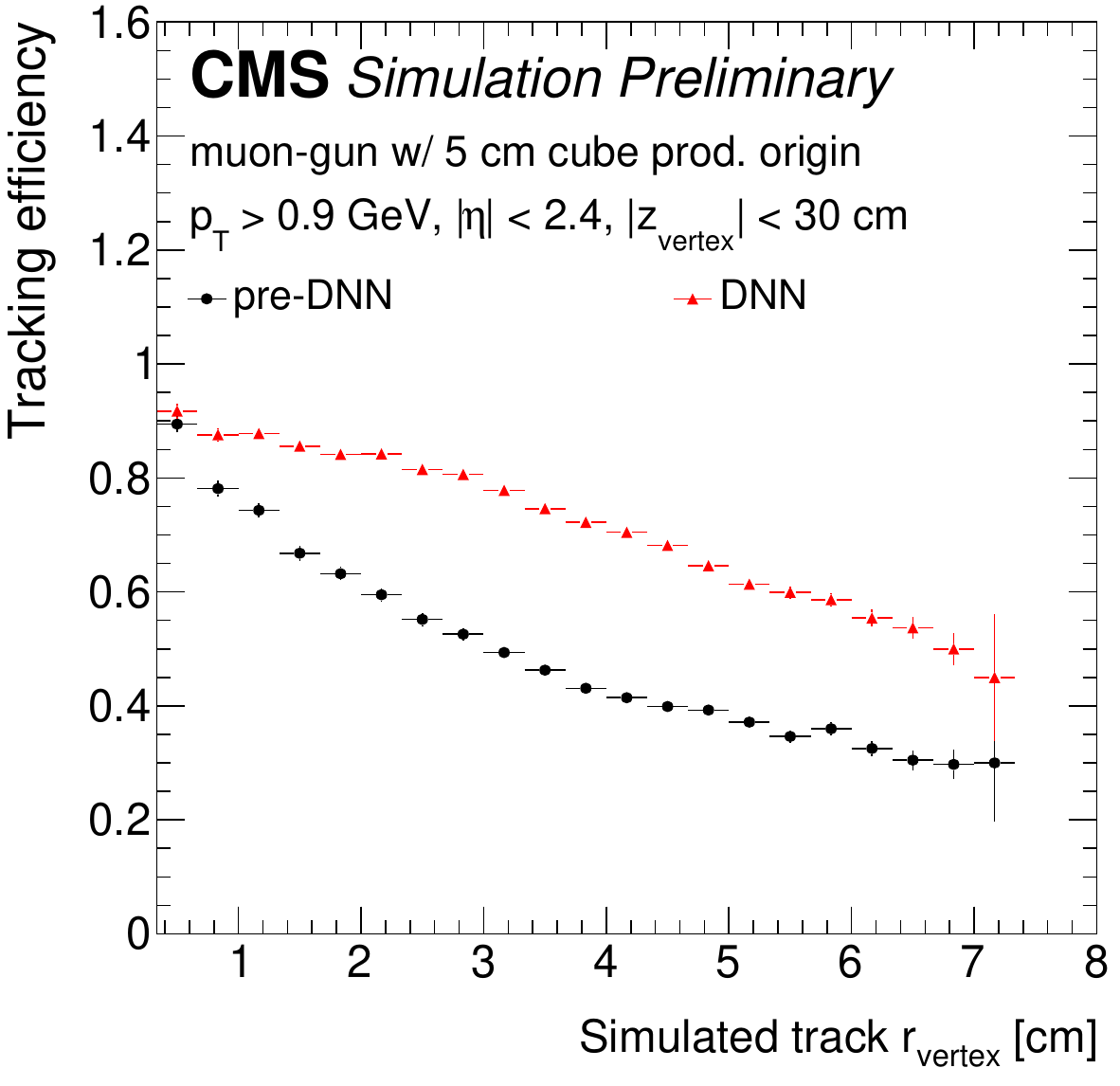}\label{fig:comparison-rvertex-cube}}
  \caption{
      The LST efficiency for all TCs is plotted as a function of $r_\mathrm{vertex}$, i.e. the distance to the production vertex measured in the plane transverse to the beamline. 
      This plot is made with 1000 $t\bar{t}$ events with HL-LHC pile-up (a) and 10,000 ``muon-cube'' events where muons are produced at points uniformly distributed across a 5 cm cube (b). 
      In both plots, it is clear that the T5-DNN recovers a significant amount of efficiency for displaced tracks. 
      In the plot made for $t\bar{t}$ events, there is a notable peak in the 18 to 20 cm $r_{vertex}$ bin. 
      This spike in efficiency is due to the detector geometry: there is less detector material in that region, so contributions from material interactions there are lower than in neighboring bins.
  }
\end{figure}

Comparisons between LST without the T5-DNN (``pre-DNN'') and with the T5-DNN were performed using 1000 $t\bar{t}$ events with HL-LHC pile-up. 
By overlaying the overall TC fake rate (Fig.~\ref{fig:comparison-fr}), it can be seen that the fake rate is reduced by approximately 40\% on average in the barrel region, where the T5 fake rate was previously the highest (Fig.~\ref{fig:lst-performance}). 
Then, by overlaying the overall TC efficiency (Fig.~\ref{fig:comparison-eff}), it is clear that, as was expected, no efficiency is lost with the T5-DNN in LST. 
Importantly, however, it can be seen that the efficiency for displaced tracks is significantly increased (Fig.~\ref{fig:comparison-rvertex-ttbar}).
The performance of LST was also compared with and without the T5-DNN using a sample of 10,000 ``muon-cube'' events, where muons are produced at points uniformly distributed in a 5 cm cube. 
With these events, it can again be seen that the efficiency for displaced tracks is greatly increased (Fig.~\ref{fig:comparison-rvertex-cube}).

\section{GNN prospects}
\label{sec:prospectus}
There is a growing number of promising results based on the TrackML dataset~\cite{cite:trackml} for an end-to-end machine learning (ML) tracking algorithm. 
Since the data is fundamentally graph-like, Graph Neural Networks (GNNs) are particularly well-suited for tracking~\cite{cite:DeZoort2021,cite:Ju2021,cite:DeZoortNature2023}. 
At the time of writing, however, there is no single-best graph-building algorithm, and no results using a full simulation of the CMS detector. 

The work shown in this paper demonstrates with full simulation that even the simplest neural network can improve upon traditional methods, so the logical next step is to try increasing the scope of the ML used in LST. 
Therefore, in future work, we plan to use LST to build the input graph for a GNN. 
If only the LSs are used to build this graph, for example, where MDs serve as nodes and LSs as edges, then the graph-building step for an LST-ML pipeline would only take 1 to 2 ms. 
Then, a GNN pipeline similar to other work (e.g.~\cite{cite:Ju2021}) could be trained on this graph to produce TCs. 
We are currently working on bringing the pixel seeds into the input graphs, such that the performance may be compared to LST as a baseline.

\section{Conclusions}
\label{sec:conclusion}
In this work, a simple two-layer DNN was trained to classify real versus fake T5s in order to improve the performance of the LST algorithm. 
Previously, T5s comprised the largest fraction of the overall efficiency, either as pT5s or T5s, as well as the bulk of the fake rate in the barrel region. 
After training DNN and incorporating the DNN into LST, we find that the DNN significantly improves the track-finding efficiency for displaced tracks while also signficantly reducing the fake rate. 
Furthermore, since the DNN gives significant improvement over pre-defined selections, we are now focusing on a more ambitious ML track-finding algorithm, where LST builds an input graph for a GNN that produces track candidates of varying sizes.


\Acknowledgements
This work was supported by the U.S. National Science Foundation under Cooperative Agreements OAC-1836650, PHY-2323298, and PHY-2121686 and grant PHY-2209443.



\end{document}